\documentstyle[12pt]{article}
\newcommand{\be}{\begin{equation}}
\newcommand{\ee}{\end{equation}}
\newcommand{\ea}{\end{array}}
\newcommand{\dis}{\displaystyle}

\begin{document}

\rightline{FTUV/94-22, IFIC/94-20}

\baselineskip 0.7cm

\begin{center}
{\Large {\bf  NEUTRINO MAGNETIC MOMENT\\

\vspace{0.2cm}

 AND THE PROCESS $\nu e \rightarrow
\nu e \gamma$}}
\end{center}

\vspace{1cm}

\begin{center}
J. Bernab\'eu$^{1)}$, S.M. Bilenky$^{1,2)}$, F.J. Botella$^{1)}$,
J. Segura$^{1)}$
\end{center}

\vspace{0.3cm}

1) Departamento de F\'{\i}sica Te\'orica, Universidad de Valencia, e IFIC,
Centro Mixto Univ. Valencia-CSIC, E-46100 Burjassot, Valencia, Spain.

2) Joint Institute for Nuclear Research, Dubna, Russia.

\vspace{3cm}

\begin{center}
{\bf Abstract.}
\end{center}

 The contribution of a neutrino magnetic moment $\mu_{\nu}$ to the
 cross section of the process
 $\nu e^{-}\rightarrow \nu e^{-} \gamma $ has been calculated and
 compared with the Standard Electroweak one. The radiative process allows
 to reach low enough values of $Q^2$ without the need to operate at very
 small energies of recoil electrons. Regions in the phase space
 which are more favourable to set bounds on $\mu_{\nu}$ are suggested.

\newpage

\section{Introduction}

\hspace{0.6cm}One of the most important problems of modern
 neutrino physics is the
investigation of neutrino properties \cite{1}: neutrino masses  and mixings,
nature of massive neutrinos (Dirac or Majorana), electromagnetic properties
of neutrinos, etc. In this paper we shall be interested in the
existence of a neutrino magnetic moment and its manifestation.

In the last few years, the interest in a magnetic moment of neutrinos was
connected in part with the solar neutrino problem.
It has been argued from time
to time that the solar neutrino flux detected in the Chlorine experiment
\cite{2}
has shown some anticorrelation with sun spot activity. Its most reasonable
explanation would involve \cite{3} a neutrino magnetic moment. Results
from Kamiokande III \cite{4} however do not indicate any time variation of
the neutrino signal. Nevertheless, the search for
a neutrino magnetic moment continues to be one of the ways to look for
 effects beyond the standard
model and efforts are worth to continue in this direction \cite{5}.

If the standard theory is extended to include the right-handed neutrino field,
the resulting Dirac neutrino with mass $m_{\nu}$ acquires a magnetic moment
 \cite{6}

\begin{equation}
\mu_{\nu}^{s} = \frac{3}{4 \sqrt{2} \pi^{2}} G_{F} m_{\nu} m \mu_{B} \simeq
3.2.10^{-19} (\frac{m_{\nu}}{eV}) \mu_{B}
\end{equation}

\noindent
where $\mu_{B} = e/2m$ is the Bohr magneton and $m$ is the electron
mass.  From the latest measurements of
the electron spectrum in $^{3}H$  $\beta$-decay \cite{7} the following
upper limit of the electron neutrino mass was obtained

\begin{equation}
m_{\nu_{e}}<7.2 eV (95\% C.L.)
\end{equation}

 It follows from (1) and (2) that the "standard" contribution (1)
 to the electron neutrino magnetic
moment is less than $2\,10^{-18} \mu_{B}$.  Such a small
upper bound cannot be reached in any present day experiment. However,
 there exist
many models beyond the standard theory in which the induced magnetic
moment of neutrinos could be many orders of magnitude bigger than $\mu_{\nu}
^{s}$ \cite{8}.

The lowest bounds on the neutrino magnetic moment come from astrophysical
arguments. If neutrinos have magnetic moments, then their coupling with an
off-shell photon $\gamma^{*}$ in a star can cause $\gamma^{*} \rightarrow
\nu + \bar{\nu}$ to occur. Once the neutrinos are produced, they will escape
carrying away energy. From the absence of such an anomalous energy loss
mechanism in red giants one finds

$$
\mu_{\nu} < 7\,10^{-11} \mu_{B}
$$

\noindent
Using the neutrino data from Supernova 1987A, there are stringent bounds which
apply for Dirac neutrinos, in order to allow the right-handed species to
escape from the supernova. One gets \cite{9}

$$
\mu_{\nu} < 10^{-12} \mu_{B}
$$

The bounds from the supernova have been questioned by Voloshin \cite{10}
if there are strong magnetic fields in the supernova. Another astrophysical
constraint comes from consideration of the luminosity before and after
stellar helium flash \cite{11} in red giants

$$
\mu_{\nu} < 3 \, 10^{-12} \mu_{B}
$$

There are laboratory bounds from terrestrial neutrino experiments. From
\mbox{$\bar{\nu_{e}} e^{-} \rightarrow \bar{\nu_{e}} e^{-}$} in reactor
 experiments \cite{12} the bound on the neutrino magnetic moment

$$
\mu_{\nu} < 2.4\,10^{-10} \mu_{B}
$$

\noindent
has been set. This limit applies to electron antineutrinos. From beam
stop LAMPF neutrino data it follows \cite{13}

$$
\mu_{\nu_{e}} < 1.1 \, 10^{-9} \mu_{B}, \quad \mu_{\nu_{\mu}}
 < 7.4 \, 10^{-10}
\mu_{B}
$$

Several new proposals \cite{14} plan to reach a much better sensitivity in the
investigation of
the $\bar{\nu}_{e}$ magnetic moment ( at the level of $10^{-11}
\mu_{B}$ ). In these experiments, the process under study to obtain information
about $\mu_{\nu}$ is that of elastic antineutrino-electron scattering at small
energies. Its sensitivity to $\mu_{\nu}$ is connected with
the fact that at low enough values of $Q^{2}$ the contribution of the
electromagnetic amplitude to the cross section of the process becomes
comparable to the contribution of the weak amplitude.
This is the case for $Q^{2} \sim MeV^{2}$ at values
$\mu_{\nu}\simeq (10^{-10}$, $10^{-11}) \mu_{B}$.
 The penetration in the region of such small
$Q^{2}$ requires, however, to measure small  energies of recoil
electrons ($\leq MeV$).

Several other transitions \cite{15} could be envisaged and have been proposed
to obtain information about the neutrino magnetic moment.
An appropriate selection of quantum numbers using nuclear transitions
to enhance the electromagnetic amplitude looks,
however, negative due to the presence of both vector and axial-vector
components in the weak amplitude, so no general enhancement of the magnetic
moment contribution relative to the weak one is found on these grounds.
Coherent neutrino-nucleus scattering keeps, as in the electron case, the
vector current contribution to both the magnetic and the weak amplitude,
but the nuclear recoil is difficult to measure ( much more difficult than
for electrons ). The strategy to get a relative enhancement of the magnetic
moment amplitude on this dynamical basis is satisfied exceptionally around
one point for electron antineutrino-electron elastic scattering, for which
there is a dynamical zero for the weak cross section \cite{16} at leading
order for $E_{\nu} = m_{e}/(4 sin^{2} \theta_{w})$ and forward electrons.

In this paper we will consider the process

\begin{equation}
\nu (\bar{\nu}) + e \rightarrow \nu (\bar{\nu}) + e + \gamma
\end{equation}

\noindent
for which there are contributions to the cross section from the weak
  interaction  as well as from the
neutrino magnetic moment. This reaction has been considered before in a
 different context \cite{17}. Even if the process (3)
 has an additional power of
$\alpha$ in the cross section relative to the elastic case, the restriction
to low recoil energies in order to reach down low values of $Q^{2}$ is a priori
not necessary. As we will see, the limit $Q^{2} = 0$ at fixed values of the
recoil energies is precisely obtained at the favourable situation of the
maximal opening angle between electron and photon in the final state. Whatever
 the
experimental limit on the total recoil energies $\nu$ could be, the inelastic
process (3) is able to lead to lower values of $Q^{2}$ than the elastic one,
as shown by the ratio $x = Q^{2} /(2 m \nu)$ varying from 1 to 0. The
argument of using low incident neutrino energies to lower the effective
contact interaction cross section of the standard theory relative to the
smoother energy dependent magnetic cross section comes here as for the
 elastic process.

The paper is organized as follows. In Section 2 we present the calculation
of the amplitudes for the process (3) and the observables.
Section 3 discusses the kinematics and the phase space details in different
variables appropriate to their experimental accessibility.
In Section 4 we analyze the behaviour of both weak and magnetic cross sections
at low $Q^{2}$ for different limiting cases: either at fixed $\nu$ or at
fixed $x$-values, by performing an analytic calculation in these limits.
General results are given in Section 5 with special emphasis in its
presentation for the inclusive distribution $d^{2} \sigma / dx d\nu$. Some
conclusions are given in Section 6.

\section{Weak and electromagnetic amplitudes}

In this Section we present shortly the results of the calculation of the
cross section for the process

\begin{equation}
\nu (l) + e (p) \rightarrow \nu (l') + e (p') + \gamma (k)
\end{equation}

The standard effective Hamiltonian of the weak interaction of neutrinos
and electrons has the form

\begin{equation}
H_{W} = \frac{G_{F}}{\sqrt{2}} \sum_{l} \bar{\nu}_{l} \gamma^{\mu}
(1 - \gamma_{5}) \nu_{l} \bar{e} \Gamma_{\mu} e + h. c.
\end{equation}

\noindent
Here

\begin{equation}
\begin{array}{lll}
\Gamma_{\mu} & = & \gamma_{\mu} [g_{L}^{(l)} \frac{\displaystyle
{1 - \gamma_{5}}}{\displaystyle{2}}
+ g_{R}^{(l)} \frac{\displaystyle{1 + \gamma_{5}}}{\displaystyle{2}}]
\end{array}
\end{equation}

\noindent
where

\begin{equation}
\begin{array}{lll}
g_{L}^{(l)} & = & - 1 + 2 sin^{2} \theta_{W} + 2 \delta_{l e}\\
\\
g^{(l)}_{R} & = & 2 sin^{2} \theta_{W}
\end{array}
\end{equation}

\noindent
and $\theta_{W}$ is the electroweak mixing angle. The term $\delta_{l e}$
in Eq. (7)  takes into account the charged current contribution to the
effective Hamiltonian in its charge-retention-form.

The invariant T-matrix element generated by the Hamiltonian (5) for the
radiative process (3) is obtained by adding the two amplitudes
 associated with the insertion of the photon in the
incoming or outgoing electron leg:

\begin{equation}
\begin{array}{rl}
T_{W}  = & \frac{G_{F}}{\sqrt{2}} e \bar{u}  (l') \gamma^{\mu}
(1 - \gamma_{5}) u (l)  \\
\\
 &  \times \bar{u} (p')  \left\{ \Gamma_{\mu}
(\hbox{\large{a}}  \varepsilon^{*} )
+ [\Gamma_{\mu} \frac{\displaystyle{\not{k} \not{\varepsilon}^{*}}}
{\displaystyle{2 (p k)}} +
\frac{\displaystyle{\not{\varepsilon}^{*} \not{k}}}{\displaystyle{2 (p' k)}}
 \Gamma_{\mu} ] \right\}
u (p)
\end{array}
\end{equation}

\noindent
where $\hbox{\large{a}}$ is the four-vector

\begin{equation}
\hbox{\large{a}}^{\alpha} = \frac{p'^{\alpha}}{(p'  k)}
- \frac{p^{\alpha}}{(p  k)}
\end{equation}

\noindent
and $\epsilon$ is the polarization vector of the photon. Let
 us notice that the use
of the Dirac equation has allowed to rewrite  the matrix element of the
process in such a way that the first term of Eq. (8) corresponds to $\gamma$-
emission by the electron charge whereas the second term is induced by the
electron magnetic moment. Such a decomposition simplifies considerably the
calculation of the cross section.

We will not enter into the details of the rather cumbersome calculations for
the cross section. Taking the appropriate sum for the neutrino spin states
(only left-handed components contribute) as well as the sum and average for
the electron spin states, one obtains from Eq. (8) the following.

\be
\begin{array}{rl}
\sum   |T_{W}|^{2}  = & 32 G_{F}^{2} e^{2} \left\{ [ - g^{2}_{L} (l p) (l' p')
- g^{2}_{R} (l' p) (lp') + g_{L} g_{R} m^{2} (ll') \right. ] \hbox{\large{a}}
^{2} \\
\\
 & +  [g_{L}^{2} (l' p') \left\{(\hbox{\large{a}}l)
 (pk) - [(\hbox{\large{a}}p) - 1] (lk) \right\} \\
\\
 & +  g_{R}^{2} (lp')  \left\{ (\hbox{\large{a}}l') (pk)
 - [(\hbox{\large{a}}p) - 1] (l' k) \right\} ]
\frac{\displaystyle{1}}{\displaystyle{(kp)}}\\
\\
& +   [g_{L}^{2} (lp) \{ (\hbox{\large{a}}l') (p'k) - [(\hbox{\large{a}} p')
 -1] (l' k) \} \\
\\
 & +  g_{R}^{2} (l' p) \{ (\hbox{\large{a}}l) (p'k) - [(\hbox{\large{a}}p')
 -1] (lk) \}  ] \frac{\displaystyle{1}}{\displaystyle{(kp')}}\\
\\
 & -  2 g_{L} g_{R} m^{2} \left. \frac{\displaystyle{(lk) (l'k)}}
{\displaystyle{(pk) (p'k)}}\right\}
\end{array}
\ee

\noindent
where $m$ is the electron mass.

We are going to take also into account  the contribution to the cross section
of the process from the diagrams with $\gamma$-exchange between neutrino and
electron vertices, due to a possible neutrino magnetic moment. The matrix
element of the electromagnetic current between initial and final neutrino
states has the form

\be
i f_{M} \sigma_{\mu \nu} q^{\nu}
\ee

\noindent
where $q = l - l' = p' + k - p$ is the momentum transfer. The coupling
$f_{M}$ at $q^{2} = 0$ is the neutrino magnetic moment $\mu_{\nu}$. We are not
going to consider a possible neutrino electric dipole moment, which is
both P-and CP-odd.

The corresponding invariant T-matrix element is given now by the amplitudes
associated to the two diagrams of Fig. 1.

\be
\begin{array}{rl}
T_{M}  = & \frac{\displaystyle{e^{2}}}{\displaystyle{q^{2}}} f_{M}
\bar{u} (l') \sigma^{\mu \nu} q_{\nu}
u (l)   \\
\\
 & \times \bar{u} (p)   \left\{ \gamma_{\mu}
(\hbox{\large{a}}  \varepsilon^{*})
+ [\gamma_{\mu} \frac{\displaystyle{\not{k} \not{\varepsilon}^{*}}}
{\displaystyle{2 (p k)}} +
\frac{\displaystyle{\not{\varepsilon}^{*} \not{k}}}{\displaystyle{2 (p' k)}}
 \gamma_{\mu} ] \right\}
u (p)
\end{array}
\ee

\noindent
with $\hbox{\large{a}}$ as given by Eq. (9). Again the two terms of Eq. (12)
correspond to \mbox{$\gamma$-emission} by the electron
 charge and magnetic moment,
respectively. In this case the neutrino vertex changes its chirality, so
for massless left handed incoming neutrinos one can obtain the
 corresponding transition probability by averaging over
initial neutrino spin states and summing over final ones. With this recipe, it
is straightforward to obtain

\be
\begin{array}{rl}
\sum |T_{M}|^{2} = & \frac{\displaystyle{32e^{4} f_{M}^{2}}}{
\displaystyle{q^{2}}} \{ (lp) (lp') \hbox{\large{a}}^{2}\\
\\
 & +  [(\hbox{\large{a}}p) (lk) - (\hbox{\large{a}}l) (pk)]
 \frac{\displaystyle{(lp')}}{\displaystyle{(kp)}}\\
\\
 &  + [(\hbox{\large{a}}p') (lk) - (\hbox{\large{a}}l)
(p'k)] \frac{\displaystyle{(lp)}}{\displaystyle{(kp')}}\\
\\
 & -  \frac{\displaystyle{(lk) (lp')}}
{\displaystyle{(kp)}} - \frac{\displaystyle{(lk) (lp)}}
{\displaystyle{(kp')}} + \frac{\displaystyle{m^{2} (kl)^{2}}}
{\displaystyle{(kp) (kp')}} \}
\end{array}
\ee

This neutrino magnetic moment contribution (13) adds incoherently to the
weak interaction result (10) as a consequence of the opposite final neutrino
helicity induced by $T_{W}$ and $T_{M}$ for massless neutrinos.

The three-body final state cross section is given, with the normalization
used for the invariant amplitudes, by

$$
d \sigma = \frac{1}{8 (lp)} \frac{1}{(2 \pi)^{5}} \delta^{4} (l + p - l' -
p' - k) \frac{d^{3}l'}{2 E'_{\nu}} \frac{d^{3} p'}{2 E'} \frac{d^{3} k}{2
E_{\gamma}} .
$$

\be
\times \sum [|T_{w}|^{2} + |T_{M}|^{2} ]
\ee

The observables of interest in terms of momenta of the recoil electron and
the emitted photon are studied in the next section.

\section{Kinematics}
 The differential cross section of the process (3) depends on 5 independent
variables. It is convenient to choose the following invariant variables

\be
\begin{array}{ll}
s & = (l + p)^{2}\\
t_{1} & = (l' - l)^{2}\\
s_{1} & = (l' + p' )^{2}\\
t_{2} & = (p - k)^{2}\\
s_{2} & = (p'+ k)^{2}
\end{array}
\ee

\noindent
for which the phase space
integral can be written as

\be
\begin{array}{c}
 \displaystyle{\int}  \frac{\displaystyle{d^{3}p'}}{\displaystyle{2 E'}}
\frac{\displaystyle{d^{3}l'}}{\displaystyle{2 E'_{\nu}}}
\frac{\displaystyle{d^{3} k}}{\displaystyle{2 E_{
\gamma}}} \delta ( p + l - p' - l' - k) =  \\
\\
=  \frac{\displaystyle{\pi}}
{\displaystyle{16 \lambda^{1/2} (s, m^{2}, 0)}} \displaystyle{\int}
\frac{\displaystyle{dt_{1} ds_{1} dt_{2}
ds_{2}}}{
\left(\displaystyle{- \Delta_{4}}\right)^{1/2}}
\end{array}
\ee

\noindent
where $\Delta_{4}$ is the 4 x 4 symmetric Gram determinant \cite{18}. The
integration domain is given by the condition $\Delta_{4}\leq 0$.

The weak and electromagnetic  squared amplitudes, as obtained in Section 2,
can be written in the form

\be
\begin{array}{l}
|T_{W}|^{2} = \frac{ \dis{f (s, t_{1}, s_{1}, t_{2}, s_{2})}}
{\dis{(s_{2} - m^{2})^{2}
(t_{2}- m^{2})^{2}}}\\
\\
|T_{M}|^{2} = \frac{\dis{g (s, t_{1}, s_{1}, t_{2}, s_{2})}}
{\dis{t_{1} (s_{2} - m^{2})^{2}
(t_{2} - m^{2})^{2}}}
\ea
\ee

\noindent
where $f$ and $g$ are third degree (or lower) polynomials of the invariants.
Fixing the other invariants, the variable $s_{1}$ corresponds to the
( unobservable ) angle between the outgoing neutrino and
electron momenta. The integration over $s_{1}$ can be performed
analytically, being $f$ and $g$ second degree polynomials in $s_{1}$. In the
Appendix we give the exact results for the triple differential cross section
 once $s_{1}$ has been integrated over, in terms of appropriate variables
(see below).

The remaining variables $t_{1}, s_{2}, t_{2}$ are observable quantities, for
which the physical region is given by the following invariant conditions:

\be
\begin{array}{l}
1) \, \,t_{1}  \leq 0; \; m^{2} \leq s_{2} \leq s, \, t_{2} \leq m^{2}\\

\\

2) \, \, G (s, t_{1}, s_{2}, 0, m^{2}, 0) \leq 0 \Rightarrow (s - s_{2})
(s - m^{2}) + st_{1} \geq 0\\

\\

3) \, \, G (s_{2}, t_{2}, 0, t_{1}, m^{2}, m^{2}) \leq 0 \Rightarrow

\left|
\begin{array}{ccccc}
0 & 1 & 1 & 1 & 1\\
1 & 0 & m^{2} & s_{2} & 0\\
1 & m^{2} & 0 & t_{1} & t_{2}\\
1 & s_{2} & t_{1} & 0 & m^{2}\\
1 & 0 & t_{2} & m^{2} & 0\\
\end{array}
\right| \geq 0
\end{array}
\ee

\noindent
with the $G$-function as defined in reference \cite{18}.

The integration over $t_{2}$ which is associated with the photon energy
$E_{\gamma}$ in the LAB frame , $t_{2}= m^{2} - 2m E_{\gamma}$ ,
 can still be
performed on analytic grounds in some cases.

Our next discussion is the translation of the physical region
(18) of the invariant variables into that for the geometrical variables in the
LAB frame:
electron-photon opening angle $\theta$, electron recoil energy $T$ and photon
energy $E_{\gamma}$, or in terms of the dimensionless variables
$x, y, \omega$ to be defined below.

The relation is given by

\be
\left\{
\begin{array}{ll}
t_{1} & \equiv - Q^{2} = 2 T \left[(m - E_{\gamma}) + E_{\gamma}
\sqrt{1 + \frac{2m}{T}}
cos \theta \right]\\
\\
s_{2} & = t_{1} + m^{2} + 2m (T + E_{\gamma})\\
\\
t_{2} & = m^{2} - 2m E_{\gamma}
\end{array} \right.
\ee

Then eqs. (18) lead to
 \mbox{$-4E_{\nu}(E_{\nu}-T-E_{\gamma})\leq t_{1} \leq 0$}
 and $E_{\nu}\geq T+E_{\gamma}$. For a given
recoil energy $T$ of the electron, the physical region in the plane $(E_{
\gamma}, \theta )$ is given by Fig. 2

There are many interesting features in Fig. 2. The line at which $Q^{2} = 0$
corresponds to the maximal opening angle

\be
Q^{2} = 0 \longleftrightarrow cos \theta = \frac{1}{\sqrt{1 + \frac{2m}{T}}}
\left( 1 - \frac{m}{E_{\gamma}}\right)
\ee

\noindent
allowed for photon energies

\be
E_{\gamma}^{0} = \frac{m}{1 + \sqrt{1 + \frac{2m}{T}}} \leq E_{\gamma}
\leq E_{\gamma}^{m} = E_{\nu} - T
\ee

For lower photon energies $0 < E_{\gamma} < E_{\gamma}^{0}$,
the maximum opening angle is 180$^{0}$ and $Q^{2}$ decreases from its
elastic scattering value $Q^{2}_{el} = 2 m T$ (at $E_{\gamma} = 0$) to
reach $Q^{2} = 0$ (at $E_{\gamma} = E_{\gamma}^{0}$). We see, therefore,
that for any values of $T$ and $E_{\gamma}$ $(T + E_{\gamma} \leq E_{\nu})$
there always exists a region of opening angles for which $Q^{2}$ is lower than
the corresponding $Q^{2}_{el}$. Furthermore, this region is found at the
 highest
allowed values of $\theta $.

Other interesting points and boundaries in Fig. 2 are the following:

 - $\theta_{1}$ is the  opening angle in the inelastic process for which
 one obtains $Q^{2} = Q^{2}_{el}$. It is given by

\be
Q^{2} = Q^{2}_{el} \longleftrightarrow cos \theta_{1} = \frac{\displaystyle{
1}}{\displaystyle{
\sqrt{1+\frac{2m}{T}}}}
\ee

 - $ \theta_{0}$ is the minimum opening angle for which $Q^{2} = 0$ is
reachable. It is given by

\be
cos \theta_{0} = [1 - \frac{m}{E_{\nu} - T} ] cos \theta_{1} \Longrightarrow
\theta_{0} > \theta_{1}
\ee

 - For the domain of the high energy photons

\be
E_{\gamma}^{1} = E_{\nu} - \frac{T}{2} (1 + \sqrt{1 + \frac{2m}{T}})
\leq E_{\gamma} \leq E_{\gamma}^{m}
\ee

\noindent
the maximum $Q^{2} = 4 E_{\nu} (E_{\nu} - T - E_{\gamma})$ corresponds to
a minimum opening  angle

\be
cos \theta = \frac{4 E_{\nu} ( E_{\nu} - T - E_{\gamma}) - 2 T (m -
E_{\gamma})}
{ 2 T E_{\gamma}\sqrt{1+\frac{2m}{T}} }
\ee
\\

It is now of interest to introduce the dimensionless variables

\be
x = Q^{2} /(2m \nu) \, , \, y = \frac{\nu}{E_{\nu}} \, , \,
\omega = \frac{E_{\gamma}}{E_{\nu}}
\ee

\noindent
with $\nu = T + E_{\gamma}$ the total energy release of the process in the
laboratory system. For fixed $x$ and $y$, the $\omega$-integration
 in the cross section, although
cumbersome, is straightforwardly made in an analytic way. We discuss some
interesting limits for the inclusive cross section in $x$ and $y$ in the
next section, in particular for its low $Q^{2}$-behaviour as a consequence
 of CVC and PCAC. First we determine the physical region in terms of these
variables, following eqs. (18):
\newpage

1)
\be
0 \leq x \leq 1
\ee

\noindent
where $Q^{2} = 0$ for $x = 0$ at  fixed $y$, whereas $E_{\gamma} = 0$
for $x = 1$

2)

\be
Q^{2} \leq 4 E_{\nu} (E_{\nu} - \nu) \Rightarrow 0 \leq y \leq (1 +
\frac{mx}{2 E_{\nu}})^{-1}
\ee

\noindent
with no threshold for the inelastic process.

3)
\be
\left\{
\begin{array}{l}
\omega^{-}  \leq \omega \leq \omega^{+} \\
\\
\omega^{\pm} =  y (1 - x) \frac{\displaystyle{1 + \frac{E_{\nu}}{m} y [1 \pm
\sqrt{ 1 + \frac{2mx}{E_{\nu} y}}]}}{\displaystyle{
1 + 2 \frac{E_{\nu}}{m} y (1 - x)}}
\end{array} \right.
\ee
\\

One notices in Eq. (29) the soft-photon limit $E_{\gamma} \rightarrow 0$
at the elastic scattering kinematics $x \rightarrow 1$. We represent
in  Fig. 3, the photon energy limits $E_{\nu} \omega^{\pm}/m$
as functions of $x$ and $\frac{E_{\nu}}{m} y$; in this
form these results are universal, independent of the incoming neutrino energy
$E_{\nu}$ except for the maximum allowed value for $\frac{E_{\nu}}{m}y$
 ( see Eq. (28) ).

In Fig. 4 we give the allowed domain of the variables $(Q^{2}, \nu)$, where
the constraint of $x$ fixed represents an straight-line
and $\nu_{0} = 2  E_{\nu}^{2} / (2 E_{\nu} + m)$.
For a given $y$, associated
for example with an experimental cut in energy release, it is possible now
to reach $Q^{2}$-values lower than $Q^{2}_{el}$ for $x < 1$. This
is nothing but a manifestation of the features discussed in Fig. 2 for the
geometrical variables. Furthermore, at fixed $x$, one can also approach $Q^{2}
\rightarrow 0$ taking $\nu \rightarrow 0$ .

\section{Low-$Q^{2}$ behaviour}

We are interested in the behaviour of both the weak and the electromagnetic
cross sections at low $Q^{2}$, with a view to enhance the second contribution
with respect to the first one. As emphasized before, it is an straightforward
, though cumbersome, matter to obtain the triple differential cross section
 in the
variables, $x, y, \omega$, as given in the Appendix; in order to check the
 results and discuss the physics of the process some limits will be
 illuminating.
 First we consider, at $y$, $\omega$ fixed, the
expansion around $x \rightarrow 0$. The weak cross section is

\be
\begin{array}{rl}
\frac{\displaystyle{d \sigma_{W}}}{\displaystyle{d x d y d \omega}}_{x<<1}
\simeq &
 \frac{\displaystyle{G^{2} m^{2}}}{\displaystyle{\pi^{2}}}\left.
\alpha \frac{\displaystyle{1}}{\displaystyle{y^{3} \omega}}
\right\{ W (y, \omega) g^{2}_{A}\\
\\
&  \left.+ \frac{\displaystyle{E_{\nu} x y}}
{\displaystyle{2m}} [V (y, \omega)
g^{2}_{V} + A (y, \omega) g^{2}_{A} +  I (y, \omega) g_{V} g_{A} ] \right\}
\end{array}
\ee

\noindent
where

\be
\begin{array}{ll}
W (y, \omega)  = & (1 - y) ( y - \omega)^{2}\\
\\
V (y, \omega)  = & (1 - y + \frac{\displaystyle{y^{2}}}{\displaystyle{2}}
) [y^{2} + \omega^{2} - \frac{
\displaystyle{2m}}{\displaystyle{E_{\nu}}}
(y - \omega) + \frac{\displaystyle{m^{2}}}
{\displaystyle{E_{\nu}^{2} y \omega}} (y - \omega)^{2} ]\\
\\
A (y, \omega)  = & (1 - y + \frac{\displaystyle{y^{2}}}{\displaystyle{2}}
) ( y^{2} + \omega^{2})\\
\\
& - \frac{\displaystyle{2m}}{\displaystyle{E_{\nu}}}
( y - \omega) [ (1 - y) \frac{\displaystyle{2y - 5 \omega}}{\displaystyle{y}}
- \frac{\displaystyle{y}}{\displaystyle{2}} (y + 2 \omega)]\\
\\
& + \frac{\displaystyle{m^{2}}}{\displaystyle{E^{2}_{\nu} y \omega}}
(y - \omega)^{2} [ (1 - y)
\frac{\displaystyle{y - 12 \omega}}{\displaystyle{y}} -
\frac{\displaystyle{y}}{\displaystyle{2}} (y + 4 \omega)]\\
\\
I (y, \omega) = & y (2 - y) (y^{2} - \omega^{2})
\end{array}
\ee

\noindent
and the couplings are

\be
g_{V} = \frac{g_{L} + g_{R}}{2}, \, g_{A} = \frac{g_{L} - g_{R}}{2}
\ee

\noindent
in terms of the chiral couplings of Eq. (7).

There are interesting features associated with this result. At $x = 0$ the
only survival term in the cross section goes like $g^{2}_{A}$. By
the use of CVC and a leptonic analogue of PCAC,
Sehgal and Weber \cite{19} reproduced this term as the
analogue of Adler's theorem for hadronic reactions. It is well known that,
due to CVC, the structure function associated with inelastic excitations
mediated by the vector current goes like $Q^{2}$ at fixed $\nu$.
So only the $g_{A}^{2}$-term can survive at $x = 0$. This term has a
 contribution for the leptonic current  proportional to the electron mass,
 hence
the global scale $m^{2}$ appearing in Eq. (30) is now understood.

Nevertheless, it is important to stress that $W(y,\omega)$ will be the dominant
 term only in a
very restricted range around $x=0$. So, for example,
this term gives a good approximation
provided $\nu>>m$ (high incoming energies) but within the restricted range
$Q^{2} < <4 m^{2}$. This is so because
the linear term in $x$, in fact, goes as
 $Q^{2}/ 4 m^{2}$.
Furthermore, the $W (y, \omega)$ dependence goes like the square
of the recoil energy of the electron. If $\nu << m$ there are high
 cancellations
in this term, seen for example when one integrates over $\omega$ at fixed $y$.
We conclude that the $x = 0$ term is only important at high incoming
energies with $\nu >> m$, but with $Q^{2} << 4m^{2}$.
Our strategy will be just the contrary, i.e., have $\nu < m$ with low
$Q^{2}$, in order to suppress the $x = 0$ $g_{A}^{2}$-term in the weak cross
 section.

The linear term in $Q^{2}$, within the bracket of Eq. (30), contains
contributions from the vector and axial couplings to electrons and their
interference. The purely vector contribution can be understood from the
Compton scattering cross section, where $y$ would be the energy of the incoming
photon and $\omega$ the energy of the outgoing photon , both normalized to
$E_{\nu}$. The Klein-Nishina formula \cite{20} for
 the cross section distribution,
when written with the appropriate variable $\omega$ instead of the
scattering angle, reads

\be
\frac{d \sigma^{\gamma \gamma}}{d \omega} = \frac{\pi \alpha^{2}}{m E_{\nu}}
\frac{1}{y^{3} \omega} [y^{2} + \omega^{2} - 2 \frac{m}{E_{\nu}} (y - \omega)
+ \frac{m^{2}}{E_{\nu}^{2} y \omega} (y - \omega)^{2} ]
\ee
\\

\noindent
which is immediately identified with the $V (y, \omega)$ term of Eq. (30).
The axial term $A (y, \omega)$ has a different behaviour and it will tend
to $V (y, \omega)$ only in the limit \mbox{$m/E_{\nu}
\rightarrow 0$}.

The cross section induced by a neutrino magnetic moment $\mu_{\nu} \not=
0$ gives, in the limit $x \rightarrow 0$.

\be
\begin{array}{ll}
\frac{\displaystyle{d \sigma_{M}}}{\displaystyle{d x dy d\omega}}_{ x < < 1}
 \simeq
& \frac{\displaystyle{\alpha^{3}}}{\displaystyle{2 m^{2}}}
 \left( \frac{\displaystyle{\mu_{\nu}}}
{\displaystyle{\mu_{B}}}\right)^{2}
\frac{\displaystyle{1}}{\displaystyle{y^{3} \omega}}
\left\{ M (y,\omega) + \frac{\displaystyle{x}}
{\displaystyle{2 y^{2} \omega}} N (y, \omega) \right\}
\end{array}
\ee
\\

\noindent
where the $(y, \omega)$-functions are

\be
\begin{array}{ll}
M (y, \omega) = & (1 - y) [(y^{2} + \omega^{2}) - 2 \frac{\displaystyle{m}}
{\displaystyle{E_{\nu}}} ( y -
\omega) + \frac{\displaystyle{m^{2}}}{\displaystyle{E_{\nu}^{2} y \omega}}
(y - \omega)^{2}]\\
\\
N (y, \omega) = & 2 y^{2} \omega (y - \omega) [(1 - y) (y + 5 \omega) + y^{2}
\omega ] \\
\\
& - \frac{\displaystyle{2m}}{\displaystyle{E_{\nu}}}
y [(1 - y ) (y^{3} - 6 y^{2} \omega + 5 y \omega^{2}
+ 6 \omega^{3})
- y^{2} \omega (y^{2} - y \omega - \omega^{2})]\\
\\
& - \frac{\displaystyle{2 m^{2}}}{\displaystyle{E_{\nu}^{2}}}
y ( y - \omega) [2 (1 - y) (3 y - 11 \omega)
+ y^{2} (y - 4 \omega)] \\
\\
& - \frac{\displaystyle{m^{3}}}{\displaystyle{E_{\nu}^{3}}} (y - \omega)^{2}
 [8 (1 - y) + 3 y^{2}]
\end{array}
\ee

The first point to be noticed in Eq. (34) is the absence of the $1/x$
singularity associated with the photon propagator in the
magnetic contribution present in the
elastic scattering cross section. This is again due to the conservation of
the electromagnetic current in the electron vertex, implying a linear
 $Q^{2}$-behaviour of the structure function, at $\nu$ fixed,  for
inelastic excitations. The leading $M (y, \omega)$ term is again, like
$V (y, \omega)$, obtainable from the Compton scattering cross section. In
fact, one can write

\be
\left.\frac{d \sigma_{M}}{d x d y d \omega} \right|_{x = 0} =
\frac{\alpha}{2 \pi} \frac{E_{\nu}}{m} (\frac{\mu_{\nu}}{\mu_{B}})^{2}
(1 - y) \frac{d \sigma^{\gamma \gamma}}{d \omega}
\ee

\noindent
with $\sigma^{\gamma \gamma}$ given by Eq. (33). Contrary to the behaviour that
we have discussed for $W (y, \omega)$ in the weak cross section, the term
$M( y, \omega)$ is not here suppressed with respect to the linear term
in $x, N (y, \omega)$, so Eq. (36) is a very good approximation to the
magnetic cross section at low energies and low values of $Q^{2}$. Taking
the ratio of cross sections at $Q^{2} = 0$, we have

\be
\begin{array}{rl}
\left.\frac{\displaystyle{d \sigma_{M}}}{\displaystyle{d \sigma_{W}}}
\right|_{x = 0}  = & (\frac{\displaystyle{\mu_{\nu}}}
{\displaystyle{\mu_{B}}})^{2}
\frac{\displaystyle{\pi^{2} \alpha^{2}}}{\displaystyle{G^{2} m^{2}}}
\frac{\displaystyle{1}}{\displaystyle{2 m g^{2}_{A} T}}  \\
\\
& \times  \left\{ \frac{\displaystyle{2 E_{\gamma} (E_{\gamma} + T)
+ T^{2}}}{\displaystyle{m T}} +
\frac{\displaystyle{m T - 2 E_{\gamma} (E_{\gamma} + T)}}
{\displaystyle{E_{\gamma} (E_{\gamma} + T)}}
\right\}
\end{array}
\ee
\\

\noindent
where the global factor in front of the bracket is a typical measure of
this ratio for the elastic scattering process at the same value of $T$. We
remind the reader that $x = 0$ is then obtained by the suitable choice of
the maximal opening angle between electron and photon. A glance at
Eq. (37) would say that the highest cross section
 ratios are obtained for the hardest
photon limit $E_{\gamma} >> T$, with values higher than the elastic ones at
will. Even more, one would say that higher neutrino energies are favoured
 in order to have hard photons but
 the discussion after Eq. (30) should have clarified that a little departure
from $x = 0$ under these conditions is enough to enhance  the
next linear term in $x$ so that the ratio (37)
becomes diluted.
To conclude, the strategy to reach low enough $Q^{2}$-values, approaching
$\theta_{max}$ at fixed $(y, \omega)$, works only in a very limited angular
range around $\theta \simeq \theta_{\max}$. Whenever the results are integrated
over a wider region of $\theta$, the
 ratio $d \sigma_{M} / d \sigma_{W}$ will be
diluted.

We can consider the approach to $Q^{2} \rightarrow 0$ following the lines
of fixed $x$ of Fig. 4. The vector contribution is in this case not
penalized due to CVC with respect to the axial
 contribution, as it was the case for $x\rightarrow 0$: the structure function
goes like $Q^{2} / \nu$ and the limit $\nu \rightarrow 0$ is not
physically forbidden for our process. It is thus of interest to study the
inclusive cross sections $d \sigma / dx dy$ and explore their
behaviour when $y \rightarrow 0$ at
fixed $x$. We can use the results of the triple differential cross sections
 given in the Appendix for the integration in
$\omega$, with the condition $\nu < < m$, and obtain

\begin{equation}
\begin{array}{ll}
\frac{\dis{d^2 \sigma_{W}}}{\dis{d x d \nu}} \simeq & \frac{\dis{4}}{\dis{3}}
 \frac{\dis{G^{2} \alpha}}{\dis{\pi^{2}}}
\frac{\dis{1}}{\dis{1-x}}\nu
\left\{x[ (g_{V}^{2} + g_{A}^{2}) - \frac{\dis{\nu}}{\dis{E_{\nu}}}
(g_{V}^2+g_{A}^2-2xg_{V}g_{A})-\frac{\dis{x}}{\dis{2}}
\frac{\dis{m \nu}}{\dis{E_{\nu}^2}} (g_{V}^2-g_{A}^2)]\right.\\
&\\
&\left.+ \frac{\dis{\nu}}{{m}}[ (\frac{\dis{17}}{\dis{10}}-2)x g_{V}^2
 +\frac{\dis{1}}{\dis{10}}(37x^2-60x+20) g^{2}_{A} ] + O(\nu^2)
\right\}
\end{array}
\ee

\noindent
for the weak cross section, whereas

\begin{equation}
\begin{array}{ll}
\frac{\dis{d^2 \sigma_{M}}}{\dis{dx d \nu}} \simeq &
\frac{ \dis{4 \alpha^{3}}}{\dis{3 m^{3}}} \left(
\frac{\dis{\mu_{\nu}}}{\dis{\mu_{B}}}\right)^{2} \frac{\dis{1}}{\dis{1-x}}
\left\{1 -\frac{\dis{\nu}}{\dis{E_{\nu}}}+\left(\frac{\dis{17}}{\dis{10}}x
-2\right)\frac{\dis{\nu}}{\dis{m}}+O(\nu^2)\right\}
\end{array}
\ee

\noindent
gives the magnetic moment cross section, which is much less sensitive to
low $x$ values. There is no need of an infrared cutoff in $\omega$ as
far as $x\neq 1$ and $\nu \neq 0$ ( see eq. (29) );
 if needed experimentally, it must be
included in this integration.

The ratio of (39) to (38) shows a very essential feature:
the most favourable sensitivity to a neutrino magnetic moment in the
inelastic process comes from the region of low excitation energy $\nu$
and, subsequently, from low $x$ values. As seen in Fig.4, lowering $\nu$
automatically lowers $Q^{2}$ and the behaviour of the structure
functions are then more favourable than for low $Q^{2}$ at fixed $\nu$.

The equations (38) and (39), valid for $x$ fixed, show the soft photon
factorization in the limit $x\rightarrow 1$. The factor in the first
 square bracket in the r.h.s. of eq. (38) is, at $x=1$, proportional to the
 weak elastic cross section up to $O(\nu^2)$;  so it is the second square
 bracket in front of $\nu /m$, once more at leading order in $\nu$. Note that
 this factor becomes $-\frac{3}{10}(g_{V}^2+g_{A}^2)$ at $x=1$. In eq. (39),
 $1-\nu/E_{\nu}$ is proportional to the elastic magnetic moment cross
 section and the remaining $-3/10$ factor of the $\nu/m$ term $x=1$ is
 the same signal of soft photon factorization as before. Finally, note
 that in eq. (38) only the last $g_{A}^2$ term survives at $x=0$, with a
 $\nu^2$ suppression due to the strong cancellations
 in the $\omega$-integration
 of $W(y,\omega)$.

\section{Results.}

\hspace{0.6cm} In this Section we present detailed numerical
 results of the weak and electromagnetic
cross section for the inelastic process (3) both for the triple
differential cross sections
$d^3\sigma_{W,M}/dTdE_{\gamma}d(cos\theta )$
and for the inclusive cross sections $d^2\sigma_{W,M}/d\nu dx$.

We have made an analysis of the ratio
$d \sigma_{M}/d \sigma_{W}$ as illustrated in figs. 5 and 6 for incoming
energies $E_{\nu} = 1 MeV$ for electron antineutrinos and $E_{\nu} = 29.79$
MeV for muon neutrinos from $\pi$-decay at rest,
respectively, using the complete expressions
without any approximations ( $T=0.2 MeV$ ).
 We give the regions in the plane $(\theta,
E_{\gamma})$ for which the cross section, when integrated from $\theta$ to
$\theta_{\max}$  ($Q^{2} = 0$)  at each $E_{\gamma}$, satisfies the following
 requirement: the ratio
$d \sigma_{M}/ d \sigma_{W}$ is 5, 4, 3 or 2 times larger than the elastic
scattering ratio for the same $T$. Even if the ratio
increases with $E_{\gamma}$ on the $Q^{2}=0$ line,
the angular width becomes more and more narrow,
as we expected from the analysis of the previous section.

Fig. 7 gives the inclusive cross section $d^2\sigma/d\nu dx$ for
electron-antineutrino scattering at \mbox{$E_{\nu} = 1 MeV$}, separating (a)
the weak contribution, (b) the magnetic moment contribution for
$\mu_{\nu}=10^{-10}\mu_{B}$, and (c) their
 ratio. The conclusion obtained in the
last section by the use of analytic limits is dramatically confirmed by
these results: the highest sensitivity is obtained for the lowest values of
$\nu$ and, by going down to low values of $x$, the sensitivity is
higher than for the elastic scattering case with $x = 1$. On the contrary,
once $\nu$ is high enough, the  sensitivity is not
improved when lowering the value of $x$. At $x = 1$ and $\nu =2E_{\nu}^2/
(2E_{\nu}+m)$
one can still see at $E_{\nu} = 1 MeV$ the residual effect of the elastic
zero present \cite{16} at $E_{\nu} = m/(4sin^2 \theta_{W})\simeq 0.51 MeV$.

Experimentally one should consider cuts both in $E_{\gamma}$ and $T$ which
 could modify our results for the inclusive cross sections which are sensitive
 to the cut in $T$ ($T^{th}$) for small $x$ and to the experimental
 threshold in $E_{\gamma}$ ($E_{\gamma}^{th}$) for
 $x\simeq 1$. However the main features remain
 the same as shown in figure 7(d), where the ratio $d\sigma_{M}/d\sigma_{W}$
 has been plotted for $E_{\nu}=1 MeV$ taking as experimental thresholds
 $T\geq T^{th}= 100 KeV$, $E_{\gamma}\geq E_{\gamma}^{th}=100 KeV$. Note that
 there is still a high sensitivity at small $\nu$ values, higher when we
 subsequently consider low $x$ values.

The general features are not highly sensitive to the incoming neutrino energy
within the range of the reactor antineutrino spectrum.
In Fig. 8  we present a similar analysis to that of
Fig. 7(c), but the cross sections have been averaged
 over a realistic \cite{21} antineutrino spectrum.
This result shows similar features as those described in the
 monoenergetic case. The only difference
 is the disappearance of the remanent
 of the elastic zero at maximum electron recoil energy, due to the average
 over incoming neutrino energy

\section{Outlook.}

\hspace{0.5cm}New neutrino physics can be
 introduced to generate a neutrino magnetic moment
as large as $10^{-10} \mu_{B}$, a magnitude which can be tested in planned
laboratory experiments on electron-antineutrino  scattering by electrons.
Furthermore this value corresponds roughly to
the scale needed to play a role in  solar
neutrino physics. The laboratory tests look for a high enough sensitivity
to the neutrino magnetic moment by lowering the accessible $Q^{2}$ to
enhance this contribution  relative to the standard weak interaction cross
section. The method based on the elastic scattering has the limitation
associated
with the cut in recoil energy needed to observe the process.

With a view to be able to reach, for a given recoil energy, lower values of
$Q^{2}$ than for the elastic process, we have studied in this paper the
weak and neutrino magnetic moment contributions to the cross section for the
inelastic radiative process $\bar{\nu} + e \rightarrow \bar{\nu} + e
+ \gamma$.

We have analyzed the inelastic process in its kinematic and dynamic behaviour
in
order to find the regions of higher sensitivity. For given recoil energies of
 both
electron and photon, the value $Q^{2} = 0$ is reachable for the highest
possible values of the opening angle between the two outgoing particles: if
$E_{\gamma} < m$, this configuration corresponds to $\theta > 90^{0}$.
The $Q^{2} = 0$ kinematic configuration is always very favourable to
enhance the neutrino magnetic moment contribution, even if at high values of
the inelastic excitation energy $\nu$ the beneficial effect of the $1/Q^{2}$
-photon propagator is lost for inelastic scattering due to CVC.
 The integration of events
around an angular region below the maximum $\theta$ dilutes, however, this
enhancement: the angular region of interest is more limited with
increasing energy of the photon. We conclude that the most interesting
situation
corresponds to the inelastic configurations $x < 1$ for low values of the
 excitation energy $\nu$. Even for the inclusive cross section $d^{2} \sigma /
dx d \nu$, this effect is clearly manifested in our results
of Figs. 7 and 8. It is understood as a suppression of the weak cross
section, Eq. (38), whereas the magnetic moment contribution has an smooth
behaviour, Eq. (39). Although absolute cross sections are small ( for instance,
$\sigma_{M}/\sigma_{W}=4.4$, $\sigma_{M}=2.7\, 10^{-47} cm^2$ for $\mu_{\nu} =
10^{-10}  \mu_{B}$ at \mbox{$E_{\nu}=1 MeV$} integrating over $\nu<0.5 MeV$ ,
 $x<0.5$ )
, the standard model contribution is suppressed in these
circumstances more strongly than in the elastic scattering case.

\vspace{3cm}
{\bf ACKNOWLEDGEMENTS}

We would like to thank  J.A. Pe\~{n}arrocha,
 L.M. Sehgal and S.K Singh, for discussions on the topic of this
paper. J. S. acknowledges the Spanish Ministry of Education and Science
for his fellowship. This work was supported in part by CICYT under Grant
AEN/93-0234.

\newpage

\appendix

\section{Appendix}

\subsection{Triple differential cross sections}

\hspace{0.5cm}We now present the exact results for the triple differential
cross
 sections in terms of the dimensionless variables (26).

The magnetic moment cross section can be written as

\be
\begin{array}{lr}
\frac{\dis{d\sigma_{M}}}{\dis{dxdyd\omega}}= & \frac{\dis{\alpha^{3}}}
{\dis{4 m^{2}}}\left(\frac{\dis{\mu_{\nu}}}
{\dis{\mu_{B}}}\right)^2\frac{\dis{1}}{\dis{(1-x)^{2}}}\frac{\dis{1}}
{\dis{(2mx+E_{\nu}y)^{2}}}\sqrt{\dis{\frac{\dis{E_{\nu}y}}
{\dis{E_{\nu}y+2mx}}}}\\
&\\
&
\times\frac{\dis{1}}{\dis{y^{3}\omega^{2}}}
[M_{0}+mM_{1}+m^{2}M_{2}+m^{3}M_{3}]
\end{array}
\ee

where

\be
\begin{array}{ll}
M_{0}=&
2E_{\nu}^{2}y^{2}(1-x)\left\{[-x(1-x)y^{2}+(6x^{2}
-6x+1)(1-y)]\omega^{3}\right.\\
&\\
&
+xy[(1-x)y^{2}-2(3x-2)(1-y)]\omega^{2}\\
&\\
&
+\left.(x^2+1)y^{2}(1-y)\omega\right\}
\\
&
\\
&
\\
M_{1}=&2yE_{\nu}\left\{-x(1-x)[(x^{2}-x+1)y^{2}+(1-y)]\omega^{3}\right.\\
&\\
&+y[-x(x-1)^{2}(x+1)y+(4x^{3}+2x^{2}-9x+2)(1-y)]\omega^{2}\\
&\\
&-(1-x)y^{2}[x(x^{2}-1)y^{2}+(x^3-13x+2)(1-y)]\omega\\
&\\
&-\left.x(x-1)^2y^3(1-y)\right\}
\\
&
\\
M_{2}=&y\left\{-x^2(1-x)y\omega^3\right.\\
&\\
& +2[x(1-x)(2x^2+x-4)y^2+(9x^2-14x+1)(1-y)]\omega^2\\
&\\
&+(1-x)y[x(x^2+x+10)(1-x)y^2-4(x^2-10x+1)(1-y)]\omega\\
&\\
&+\left.2y^2(x-1)^2[-x(1-x)y^2+(x^2-6x+1)(1-y)]\right\}
\\
&
\\
M_{3}=&\frac{\dis{x}}{\dis{E_{\nu}}}[-(1-x)(x+3)y^2-8(1-y)](\omega-y (1-x))^2
\end{array}
\ee

\newpage

The weak cross section reads

\be
\begin{array}{rl}
\frac{\dis{d\sigma_{W}}}{\dis{dxdyd\omega}}= &
\frac{\dis{G^2 m}}{\dis{4\pi^2}}\alpha
\frac{\dis{1}}{\dis{(1-x)^{2}}}\frac{\dis{1}}
{\dis{(2mx+E_{\nu}y)^{2}}}\sqrt{\dis{\frac{\dis{E_{\nu}y}}
{\dis{E_{\nu}y+2mx}}}}\\
&\\
&
\times\frac{\dis{1}}{\dis{y^{2}\omega^{2}}}
[W_{0}+mW_{1}+m^2 W_{2}+m^{3}W_{3}+m^{4}W_{4}]
\end{array}
\ee

where

$$
\begin{array}{ll}
W_{0}=&x(1-x)E_{\nu}^3y^2\left\{y^2(x^2+1)[-y(2-y)(g_{V}-g_{A})^2+
2(g_{V}^2+g_{A}^2)]\omega\right.\\
&\\
&-2xy[(xy^2+2(3x-2)(1-y))(g_{V}^2+g_{A}^2)+2x(2-y)g_{V}g_{A}]\omega^2\\
&\\
&+[((2x^2-2x+1)y^2+2(6x^2-6x+1)(1-y))(g_{V}^2+g_{A}^2)\\
&\\
&\left.+2(2x-1)y(2-y)g_{V}g_{A}
]\omega^3\right\}
\\
&
\\
W_{1}=&E_{\nu}^2 y\left\{(x-1)^2x^2y^3[y(2-y)(g_{V}-g_{A})^2
-2(g_{V}^2+g_{A}^2)]\right.\\
&\\
&+2(1-x)y^2[xy^2((x^3+4x-1)g_{V}^2+(x^3+4x+1)g_{A}^2)\\
&\\
&-(1-y)(x(x^3-13x+2)g_{V}^2+(x^4-15x^2+6x-2)g_{A}^2)\\
&\\
&+2x^2(x^2+4)y(2-y)g_{V}g_{A}]\omega\\
&\\
&-y[xy^2((2x^4-10x^3+6x^2+5x-2)g_{V}^2\\
&\\
&+(2x^4-10x^3+2x^2+9x-2)g_{A}^2)\\
&\\
&-2(1-y)(x(4x^3+2x^2-9x+2)g_{V}^2\\
&\\
&+(4x^4+14x^3-37x^2+22x-4)g_{A}^2)\\
&\\
& -2(4x^3-4x^2+4x-5)x^2y(2-y)g_{V}g_{A}]\omega^2\\
&\\
&+2(1-x)[-xy^2(x(x^2-x-1)g_{V}^2+(x^3-x^2-3x+2)g_{A}^2)\\
&\\
&\left.-(1-y)(x^2g_{V}^2+(-11x^2+12x-2)g_{A}^2)]\omega^3\right\}
\end{array}
$$
\\
\be
\begin{array}{ll}
W_{2}=&E_{\nu}x\left\{(x-1)^2y^3[-y^2((2x^2+2x-1)g_{V}^2+(2x^2+2x+1)g_{A}^2)
\right.\\
&\\
&+2(1-y)(x^2-6x+1)(g_{V}^2+g_{A}^2)-4y(2-y)(x+1)xg_{V}g_{A}]\\
&\\
&+(1-x)y^2[y^2((3x^4+3x^2+2x-2)g_{V}^2+(3x^4-x^2+26x-2)g_{A}^2)\\
&\\
&-4(1-y)((x^2-10x+1)g_{V}^2+2(x^2-6x+1)g_{A}^2)\\
&\\
& +8y(2-y)(3x+1)xg_{V}g_{A}]\omega\\
&\\
& +y[y^2((4x^4-6x^3-2x^2+1)g_{V}^2+(4x^4-10x^3+30x^2-36x+7)g_{A}^2)\\
&\\
&+2(1-y)((9x^2-14x+1)g_{V}^2+(13x^2-22x+5)g_{A}^2)\\
&\\
&+4y(2-y)(4x^2-5x-1)xg_{V}g_{A}]\omega^2\\
&\\
&\left.+(1-x)[y^2(3x^2g_{V}^2+(-x^2+4x-4)g_{A}^2)-4(1-y)g_{A}^2]\omega^3\right\}
\\
&
\\
W_{3}=&x^2\left\{(1-x)^2y^2[y^2((1-x)(3x+1)g_{V}^2-(3x^2-2x+7)g_{A}^2)
\right.\\
&\\
&-8(1-y)(g_{V}^2+g_{A}^2)-8y(2-y)xg_{V}g_{A}]\\
&\\
&+2(1-x)y[y^2((3x^2-6x-1)g_{V}^2+(2x^2+4x+6)g_{A}^2)\\
&\\
&+8(1-y)(g_{V}^2+g_{A}^2)+8y(2-y)xg_{V}g_{A}]\omega\\
&\\
&+[-y^2((11x^2-10x-1)g_{V}^2+(-9x^2+14x+3)g_{A}^2)\\
&\\
&-8(1-y)(g_{V}^2+g_{A}^2)-8y(2-y)xg_{V}g_{A}]\omega^2\\
&\\
&\left.+2(1-x)yg_{A}^2\omega^3\right\}
\\
&
\\
W_{4}=&\frac{\dis{4x^3y}}{\dis{E_{\nu}}}[\omega-y(1-x)]^2(g_{V}^2-g_{A}^2)
\end{array}
\ee

\subsection{Soft-photon limit}

\hspace{0.5cm}From the expressions for the
 triple differential cross sections given
 above we can now perform the integration over $\omega$
 around $x\simeq 1$; the result both for the weak and the magnetic cross
 section reads

\begin{equation}
\frac{\dis{d^2\sigma}}{\dis{dxdy}}=\frac{\dis{\alpha}}{\dis{\pi}}
\left[\frac{\dis{d\sigma}}{\dis{dy}}\right]_{el}\frac{\dis{1}}{\dis
{1-x}}[F(z)+O(1-x)]
\end{equation}

\noindent
where $\left[\frac{\dis{d\sigma}}{\dis{dy}}\right]_{el}$ is the corresponding
 elastic scattering cross section, taking $y=T/E_{\nu}$, and

\begin{equation}
F(z)=\left[-2+\frac{\dis{1}}{\dis{z}}ln\left(\frac{\dis{1+z}}{\dis{
1-z}}\right) \right]=2 \sum_{n=1}^{\infty}\frac{\dis{z^{2n}}}{\dis{2n+1}}
\end{equation}

\noindent
where
\begin{equation}
z=\frac{\dis{\sqrt{\nu^2+2m\nu}}}{\dis{m+\nu}}<1
\end{equation}
\\

{}From this result it is easy to check the $\nu/m\rightarrow 0$ limit in
 eqs. (38) and (39) for $x\rightarrow 1$.
Notice the logarithmic infrared divergence of
 $d\sigma/dy$.

\newpage
\begin{center}
{\large {\bf Figure Captions}}
\end{center}

\begin{itemize}
     \item{{\bf Fig. 1)} Electromagnetic interaction Feynman
 diagrams for the process
$\nu e^- \rightarrow \nu e^- \gamma$.}
     \item{{\bf Fig. 2)} Physical region in the plane $(E_{\gamma},\theta)$
 for a fixed $T$-value.}
     \item{{\bf Fig. 3)} Photon energy limits $E_{\gamma}^{\pm}/m$ as functions
  of $x$ and $\nu/m$; these limits are independent of incoming neutrino
  energy except for the maximum value of $\nu$,
 $\nu\leq 2E_{\nu}^2/(2E_{\nu}^2+mx)$.}
     \item{{\bf Fig. 4)} Allowed domain for the variables $(Q^2,\nu)$. The
 straight lines passing through the origin represent fixed $x$ values. $x=1$
 ( upper line ) corresponds to the soft photon limit.}
     \item{{\bf Fig. 5)} Regions in the plane $(\theta,E_{\gamma})$
 where the $\bar{\nu}_{e}$
 radiative cross sections, when integrated from $\theta$ to $\theta_{max}$
 $(Q^2=0)$, satisfy that the ratio $d\sigma_{M}/d\sigma_{W}$ is 5,4,3 or 2
 times larger than the elastic ratio at the same $T$-value $(T=0.2MeV)$.
 The solid line represents the $Q^2=0$ curve. In this figure $E_{\nu}=1 MeV$.}
     \item{{\bf Fig. 6} Same as figure 5 but for muon antineutrinos from
 pion decay at rest $(E_{\nu}=29.79 MeV)$. In this figure $T=0.2MeV$.}
     \item{{\bf Fig. 7(a)} Inclusive weak cross section $d^2\sigma_{W}
/dxd\nu$ for electron antineutrinos. The physical region is bounded by
 $0\leq x\leq 1$ and $0\leq \nu \leq 2E_{\nu}^2/(2E_{\nu}+mx)$; the flat region
 on the right is unphysical. The decimal logarithm of the cross section in
 $10^{-45} cm^2/MeV$ units is represented.}
     \item{{\bf Fig. 7(b)} Same as 7(a) but for the magnetic moment cross
 section with \mbox{$\mu_{\nu}=10^{-10}\mu_{B}$}.}
     \item{{\bf Fig. 7(c)} Same as 7(a) for the ratio $d^2\sigma_{M}
/d^2\sigma_{W}$.}
      \item{{\bf Fig. 7(d)} Same as 7(c) including thresholds both for
 $E_{\gamma}$ and $T$ ( $E_{\gamma}^{th}=100 KeV$ and $T^{th}=100 KeV$).
 The flat regions ($d^2\sigma_{M}/d^2\sigma_{W}=0$) are unphysical.}
     \item{{\bf Fig. 8} Same as figure 7(c) but averaged
 over a realistic reactor antineutrino spectrum.}
 \end{itemize}

\end{document}